\documentclass[a4paper]{article}

\usepackage{url}
\usepackage{times}
\usepackage{verbatim}
\usepackage[T1]{fontenc}
\usepackage[english]{babel}
\usepackage{amsmath}
\usepackage{amssymb}
\usepackage[dvipdf]{graphicx}
\usepackage{color}
\usepackage[utf8]{inputenc}
\usepackage{tikz}
\usepackage{graphicx}
\usetikzlibrary{positioning}
\usepackage{epsfig}
\usepackage{float}
\usepackage[small]{caption}
\usepackage{subfig}
\usepackage{mathtools}
\mathtoolsset{showonlyrefs}
\usepackage{natbib}
\graphicspath{{.}}

\def\R{\mathbb{R}}

\title{On the modeling assumptions of Historical Simulation for Value-at-Risk}

\author{
    Björn Löfdahl Grelsson\footnote{bjorn.lofdahl@seb.se}
}

\date{\today}

\begin{document}
\maketitle

\begin{abstract}
Historical Simulation (HS) and its extensions form a popular class of methods for estimating Value-at-Risk for portfolios of financial assets based on historical data. In this note, we seek to unify several ideas and models from throughout the literature into a single modeling framework. By explicitly defining a parametric model form for the asset returns and extracting the realized increments of the driving innovation process from historical data, we are able to reproduce the Historical Simulation, filtered Historical Simulation, and displaced Historical Simulation methods. This shows beyond a doubt that these methods need more underlying assumptions than what is often alluded to.  
\end{abstract}

\vspace{4mm}

% ════════════════════════════════════════════════════════════
\section{Introduction}\label{sec:intro}

Historical Simulation (HS) and its extensions form a popular class of methods for estimating Value-at-Risk (VaR) for portfolios of financial assets based on observed historical prices. It has been extensively studied in the literature; see e.g. \cite{Alexander2008}, \cite{Meucci2005}, \cite{Hult2012}, and others. In the method's simplest form, the asset price in each scenario is determined by considering the market state today, and adding a risk factor perturbation that corresponds to a daily risk factor change between two days in the risk factor's historical sample. \cite{Hult2012} note that relative shifts may be assumed to be representative of future returns of stock prices, i.e., daily risk factor shifts are calculated as
\begin{equation}\label{eq:rel_shift}
    r_k^{\mathrm{rel}} = \frac{S_k}{S_{k-1}} - 1,
\end{equation}
where $r_k$ and $S_k$ denote the risk factor shift and asset price, respectively, for day $k$. For the relative shifts \eqref{eq:rel_shift} to be meaningful and well defined, the denominator must be positive. If the denominator is zero, the shift is not well defined. If the denominator is negative, the risk factor shift will have the opposite sign of the actual risk factor move for the two consecutive historical days. If the denominator is very small, but positive, and the numerator is of ``average magnitude'', the daily shift can be arbitrarily large. Hult et al. instead propose that absolute shifts may be viewed as good representatives for future returns of interest rates, i.e., the daily risk factor shifts are calculated as
\begin{equation}\label{eq:abs_shift}
    r_k^{\mathrm{abs}} = S_k - S_{k-1}.
\end{equation}
They also note that if the generated sample of returns can be viewed as samples from i.i.d. random variables, then standard statistical techniques can be used to investigate the probability distribution of future portfolio values. However, they leave it up to the reader to verify these assumptions. \cite{Meucci2005} formalizes the concept of market invariants as quantities whose distribution is stable over time. He gives several explicit examples: log-returns for equities, yield changes for rates, and others. 

The simple fact that the properties of the price or rate process itself influence the way shifts are calculated should give the reader a hint that there is more to this problem than meets the eye, and that there may be some hidden assumptions behind the choice of shift transform. In fact, the supposedly model-free assumption of the method has already been discussed, directly or indirectly, by several authors. \cite{Alexander2008} proposes that the main advantage of historical VaR is that there is no need to make an assumption about the parametric form of the distribution of the risk factor returns. Instead, the dynamic evolution and the dependencies of the risk factors are inferred directly from historical observations. However, Alexander also notes that asset volatility may fluctuate over time, and suggests that the historical returns could in fact be scaled according to the GARCH or exponentially weighted moving average (EWMA) volatility models, following \cite{Bollerslev1986}, \cite{duffie1997overview}, and \cite{hull1998value}. The methodology is designed to weight historical returns so that their volatility is adjusted to the current volatility.

\cite{Pritsker2006} states that the principle advantage of HS is that it is in some sense nonparametric, because it does not make any assumptions about the shape of the distribution of the risk factors that affect the portfolio's value. Because the distribution of risk factors, such as asset returns, is often fat-tailed, historical simulation might be an improvement over other VaR methods which assume that the risk factors are normally distributed. Pritsker further notes that the main insight of the filtered HS (FHS) \citep{BaroneAdesi1999} method is that it is possible to capture conditional hetero-skedasticity in the data and still be somewhat unrestrictive about the shape of the distribution of the factors returns. He then states that the method appears to combine the best elements of conditional volatility models with the best elements of the historical simulation method.

In this paper, we seek to unify several ideas and models from throughout the literature into a single framework. We show that the standard shift rules in Historical Simulation follow naturally from two special cases of a rather general volatility scaling function, and that acknowledging this opens a natural path to local and stochastic volatility extensions. While the individual models considered here are well established, the explicit identification of a common innovation-extraction mechanism, and the demonstration that the choice of shift rule is equivalent to a choice of volatility function, has, to the best of our knowledge, not been presented in this form. By reducing all three methods to instances of the same volatility-scaling formula, it becomes clear that the apparent variety of shift rules in the literature reflects not different methodologies, but different modeling assumptions about the same underlying quantities — the local and stochastic volatility of the price process. In all cases, the adequacy of the chosen model can be assessed by testing the extracted innovation increments for serial dependence and heteroskedasticity. The framework further clarifies the distinction between VaR and stressed VaR in the presence of stochastic volatility: the former requires rescaling historical returns by the ratio of current to historical volatility, while the latter should not, since the purpose is precisely to preserve the elevated volatility of the historical sample.

The remainder of the paper is organized as follows: Section \ref{sec:framework} starts from a rather simple time-discrete stochastic process with local volatility, and derives the standard shift rules from two special cases.  Section \ref{sec:lsv} builds on these ideas and extends the model with a stochastic volatility process, outlining how this additional assumption is equivalent to more advanced semi-parametric methods for HS, such as FHS.

% ════════════════════════════════════════════════════════════
\section{Historical simulation as outcomes of an innovation process}\label{sec:framework}

Consider a discrete-time stochastic process $S$ with increments given by
\begin{equation}\label{eq:general_dynamics}
    \Delta S_k = S_k - S_{k-1} = \gamma(S_{k-1})\,\Delta W_k,
\end{equation}
where $\gamma$ is the volatility (diffusion) function of the process, and $W$ denotes an innovation process with i.i.d. increments. A reader familiar with Ito-style diffusion processes may, if it makes her comfortable, think about $W$ as a discrete-time analogue of the standard Wiener process, although we do not need to impose any restrictions with respect to its distribution for the purpose of our analysis. Under the bold, to put it mildly, assumption that the function $\gamma$ is known, we can easily extract the innovation process increments as
\begin{equation}\label{eq:extract_innovations}
    \Delta W_k = \frac{S_k - S_{k-1}}{\gamma(S_{k-1})}.
\end{equation}
From the empirical increments \eqref{eq:extract_innovations}, we create simulated market states $\tilde{S}_k$, $k=1,\ldots,N$ by advancing the process $S$ one time step from the base market state $S_0$. That is, the simulated scenarios are given by

\begin{equation}\label{eq:simulated_scenarios}
    \tilde{S}_k = S_0 + \gamma(S_0)\,\Delta W_k
               = S_0 + \frac{\gamma(S_0)}{\gamma(S_{k-1})}\,(S_k - S_{k-1}).
\end{equation}
\cite{lichtner2019choose} derives the same formula \eqref{eq:simulated_scenarios} by introducing the notion of generalized returns, although he does not explicitly link it to an underlying innovation process, at least to our understanding. Assuming an arithmetic Brownian motion (ABM), e.g. Hull-White type model or similar, with\ $\gamma(x) = \sigma \in \R^+$, we get
\begin{equation}\label{eq:abm}
    \tilde{S}_k = S_0 + \frac{\sigma}{\sigma}\,(S_k - S_{k-1})
               = S_0 + r_k^{\mathrm{abs}},
\end{equation}
which corresponds directly to the absolute shift rule. Assuming instead a geometric Brownian motion (GBM), e.g. Black-Scholes type model or similar, with $\gamma(x) = \sigma x,\; \sigma \in \R^+$, we obtain
\begin{equation}\label{eq:gbm}
    \tilde{S}_k = S_0 + \frac{\sigma \cdot S_0}{\sigma \cdot S_{k-1}}\,(S_k - S_{k-1})
               = S_0 + S_0 \cdot r_k^{\mathrm{rel}},
\end{equation}
which corresponds directly to the relative shift rule. From these two simple examples, we can easily see that the standard shift rules for HS follow from the two simple special cases \eqref{eq:abm}-\eqref{eq:gbm} of a much more general model, and, that historical simulation, while being parameter free, is not at all model free. The assumptions are simply hidden under a mask of simplicity.

For the simulation \eqref{eq:simulated_scenarios} to be statistically sound, the innovation process increments \eqref{eq:extract_innovations} should ideally be i.i.d. If $\gamma$ is misspecified, the resulting innovation series may exhibit non-stationarity, which would invalidate the core assumption underlying any resampling procedure. In practice, the volatility function $\gamma$ is of course unknown and can only be estimated based on historical returns.

\subsection{On the relation to displaced historical simulation}\label{sec:dhs}
The so called displaced Historical Simulation method introduced by \cite{fries2017displaced} considers the special case where the innovation process is standard Wiener, and the volatility function is given by $\gamma(x) = \Big((1-  |\alpha| )x + \alpha \beta\Big)\sigma$. For any $-1\leq\alpha\leq1$, the model is a mixture of a normal and a log-normal process, while for $\alpha \in \{-1,0,1\}$ the model collapses to the standard normal and log-normal cases, respectively. Interestingly, this mixture model leads to a hybrid shift rule. Using our notation, the authors arrive at

\begin{equation}\label{eq:mixed_shifts}
\tilde S_k = S_0 + \alpha(S_{k-1}) S_0 r_{k}^{rel} + \left(1-\alpha(S_{k-1})\right) r_{k}^{abs},
\end{equation}
where $\alpha(x) = \frac{x}{x+a}$, and $a=\frac{\alpha}{1-|\alpha|}\beta$. This too, turns out to be a special case of our model. The function $\alpha$ determines how much emphasis are put on absolute and relative shifts, respectively. If $\alpha=0$, then we perform a pure absolute shift. If $\alpha=1$, we perform a pure relative shift. Note that, for any $\alpha \in [0,1]$, the method is historically consistent, in the sense that applying it to a one-day change from, say, yesterday till today, gives us back today's price:
\begin{equation}
  \tilde{S}_k = S_{k-1} + \alpha S_{k-1} r_{k}^{rel} + \left(1-\alpha\right) r_{k}^{abs} = S_k.
\end{equation}

Now, comparing \eqref{eq:mixed_shifts} with the volatility scaling model \eqref{eq:simulated_scenarios}, we identify
\begin{equation}
  S_0 + \frac{\gamma\left(S_0\right)}{\gamma\left(S_{k-1}\right)} \left(S_k - S_{k-1}\right) = S_0 + \alpha S_0 r_{k}^{rel} + \left(1-\alpha\right) r_{k}^{abs},
\end{equation}
and obtain
\begin{equation}
  \frac{\gamma\left(S_0\right)}{\gamma\left(S_{k-1}\right)} = 1 + \alpha \frac{S_0 - S_{k-1}}{S_{k-1}}.
\end{equation}
Solving for $\alpha$ explicitly,
\begin{equation}
  \alpha\left(S_0, S_{k-1}\right) = \frac{S_{k-1}}{S_0 - S_{k-1}} \left(\frac{\gamma\left(S_0\right)}{\gamma\left(S_{k-1}\right)} - 1\right),
\end{equation}
for $S_0 \neq S_{k-1}$. We see that modeling the function $\alpha$ is equivalent to modeling the ratio of the volatilities. It then becomes a modeling choice whether to model the volatility directly, or instead to model the interpolation function $\alpha$. We prefer the former, due to the fact that the assumptions on the underlying process become explicit rather than implicit.

\section{Extension to local-stochastic volatility}\label{sec:lsv}

We now consider the extension where the stochastic process $S$ has discrete-time increments given by
\begin{equation}\label{eq:lsv_general}
    \Delta S_k = S_k - S_{k-1} = \mu(S_{k-1}) + \gamma(S_{k-1},\, v_{k-1})\,\Delta W_k,
\end{equation}
where $\mu$ denotes the drift of the process, $\gamma$ is the volatility function of the process, $v$ denotes a stochastic volatility process, and $\Delta W_k$ denotes an increment of the innovation process. For the remainder of this paper, to make our argument clear without introducing unnecessary complexity, we will only consider the case with zero drift and a multiplicative form for $\gamma$, i.e.,
\begin{equation}\label{eq:lsv}
    \Delta S_k = S_k - S_{k-1} = v_{k-1}\,\gamma(S_{k-1}) \Delta W_k.
\end{equation}
The zero drift assumption is reasonable over short time horizons, such as one single day. Following the same procedure as in Section~\ref{sec:framework}, the innovation process increments are given by
\begin{equation}\label{eq:extract_innovations_svar}
    \Delta W_k = \frac{S_k - S_{k-1}}{v_{k-1}\gamma(S_{k-1})},
\end{equation}
and the simulated prices, analogous to \eqref{eq:simulated_scenarios}, are given by
\begin{equation}\label{eq:lsv_scenario}
    \tilde{S}_k = S_0 + v_0\,\gamma(S_0)\,\Delta W_k
               = S_0 + \frac{v_0\,\gamma(S_0)}{v_{k-1}\,\gamma(S_{k-1})}\,(S_k - S_{k-1}),
\end{equation}
that is, price changes are now scaled not only by relative local volatility, but also by the ratio of the stochastic volatilities $\frac{v_0}{v_{k-1}}$.

Assuming an extended Black-Scholes type model with $\gamma(x) = \sigma x,\; \sigma \in \R^+$, such as the Heston or Bates stochastic volatility models or the GARCH model, we obtain
\begin{equation}\label{eq:fhs_gbm}
    \tilde{S}_k = S_0 + S_0 \cdot \frac{v_0}{v_{k-1}} \frac{S_k - S_{k-1}}{S_{k-1}},
\end{equation}
that is, we scale the observed relative return with the ratio of volatilities. This corresponds directly to the methodology of filtered HS due to \cite{BaroneAdesi1999}. Assuming a GARCH-type model with
\begin{align}\label{eq:garch}
    r_k &= \frac{\Delta S_k}{\gamma(S_{k-1})} = v_{k-1}\Delta W_k, \\
    v_k^2 &= a_0 + a_1r_{k-1}^2 + b_1v_{k-1}^2,
\end{align}
we immediately arrive at a generalization of a model outlined by \cite{Pritsker2006}. The EWMA model follows as a special case, with
\begin{equation}\label{eq:ewma}
    v_k^2 = (1-\lambda)\,r_{k-1}^2 + \lambda\,v_{k-1}^2
          = (1-\lambda)\sum_{i=1}^{k} \lambda^{i-1}\,r_{k-i}^2,
\end{equation}
where $\lambda$ is a smoothing parameter. We immediately see that the GBM model with $\gamma(x) = \sigma x$ assumes relative shifts for the volatility estimation, and that the ABM model with $\gamma(x) = \sigma$ assumes absolute shifts. We further note that in both cases, the parameter $\sigma \in \R^+$ will vanish when we consider the ratio $\frac{v_k}{v_n}$ for any $k$ and $n$, i.e., the model remains parameter free with respect to the local volatility for these special cases.

We note that, assuming a parametric form $\gamma(\cdot) = \gamma(\cdot\,;\, \theta)$ for the local volatility function, the parameters $\theta$ and the GARCH coefficients can in principle be estimated jointly via quasi-maximum likelihood \citep{BollerslevWooldridge1992}. Under the model \eqref{eq:garch}, the conditional quasi-log-likelihood takes the form
\begin{equation}\label{eq:qmle}
    \ell(\theta, a_0, a_1, b_1) = -\frac{1}{2} \sum_{k=1}^{N} \left[\log v_{k-1}^2 + \log \gamma^2(S_{k-1};\, \theta) + \frac{(\Delta S_k)^2}{v_{k-1}^2\, \gamma^2(S_{k-1};\, \theta)}\right],
\end{equation}
where
\begin{align}
    r_k &= \frac{\Delta S_k}{\gamma(S_{k-1};\, \theta)}, \\
    v_k^2 &= a_0 + a_1 r_{k-1}^2 + b_1 v_{k-1}^2.
\end{align}
A full empirical investigation of this approach is outside the scope of this note and is left to future work.

We end this section with a note on the concept of stressed VaR. The purpose of stressed Historical VaR is to understand what your portfolio VaR would be if the market shifted to a more volatile regime. In the context of HS, this is often calculated by choosing the historical sample from a period with elevated volatility, such as the Great Financial Crisis, the Euro crisis, the outbreak of the Ukraine War, and so on. According to \cite{lichtner2019choose}, choosing an appropriate return model for historical returns is of the utmost importance to be able to accurately estimate stressed VaR. We note that model \eqref{eq:lsv} is well suited to this purpose if we slightly alter the simulation equation \eqref{eq:lsv_scenario}. To preserve the high volatility of the historical sample, we should consider the simulation scheme
\begin{align}\label{eq:lsv_stressed_scenario}
    \bar{S}_k &= S_0 + v_{k-1}\,\gamma(S_0)\,\Delta W_k\\
               &= S_0 + \frac{v_{k-1}\,\gamma(S_0)}{v_{k-1}\,\gamma(S_{k-1})}\,(S_k - S_{k-1})\\
               &= S_0 + \frac{\gamma(S_0)}{\gamma(S_{k-1})}\,(S_k - S_{k-1}),
\end{align}
that is, the simulation equation reduces to the case without a stochastic volatility component. However, for the purposes of fitting a model to data, one should carefully consider whether a joint estimation of local and stochastic volatility is still preferable, since the same calibrated model can then be used to estimate both VaR and stressed VaR. 

% ════════════════════════════════════════════════════════════
\section{Conclusions}\label{sec:conclusion}
We have shown that several prominent methods for calculating VaR, namely Historical Simulation, filtered Historical Simulation, and displaced Historical Simulation, can be seen as special cases of a more general volatility scaling method. By explicitly defining a parametric model form for the asset returns and extracting the historical increments of the driving innovation process, it becomes clear that the choice of shift rule is equivalent to a choice of local volatility function, and that the adequacy of this choice can be assessed by testing the extracted innovations for serial dependence and heteroskedasticity. The framework further provides a natural distinction between VaR and stressed VaR: the former requires rescaling by the ratio of current to historical stochastic volatility, while the latter should not. We argue that the practical implication of these results should be that risk managers make their dynamics choice explicit and test its adequacy, rather than hiding behind the model-free illusion that is Historical Simulation.

\bibliography{model_free_myth}

@book{Alexander2008,
  author    = {Carol Alexander},
  title     = {Market Risk Analysis, Volume {IV}: Value-at-Risk Models},
  publisher = {John Wiley \& Sons},
  year      = {2008}
}

@book{Hult2012,
  author    = {Henrik Hult and Filip Lindskog and Ola Hammarlid and Carl Johan Rehn},
  title     = {Risk and Portfolio Analysis: Principles and Methods},
  publisher = {Springer},
  year      = {2012}
}

@book{Meucci2005,
  author    = {Attilio Meucci},
  title     = {Risk and Asset Allocation},
  publisher = {Springer},
  year      = {2005}
}

@article{Pritsker2006,
  author  = {Matt Pritsker},
  title   = {The Hidden Dangers of Historical Simulation},
  journal = {Journal of Banking \& Finance},
  volume  = {30},
  number  = {2},
  pages   = {561--582},
  year    = {2006}
}

@article{duffie1997overview,
  title={An overview of value at risk},
  author={Duffie, Darrell and Pan, Jun and others},
  journal={Journal of derivatives},
  volume={4},
  number={3},
  pages={7--49},
  year={1997}
}

@article{fries2017displaced,
  title={Displaced relative changes in historical simulation: Application to risk measures of interest rates with phases of negative rates},
  author={Fries, Christian P and Nigbur, Tobias and Seeger, Norman},
  journal={Journal of Empirical Finance},
  volume={42},
  pages={175--198},
  year={2017},
  publisher={Elsevier}
}

@article{lichtner2019choose,
  title={How to choose the return model for market risk? Getting towards a right magnitude of stressed VaR},
  author={Lichtner, Mark},
  journal={Quantitative Finance},
  volume={19},
  number={8},
  pages={1391--1407},
  year={2019},
  publisher={Taylor \& Francis}
}

@article{hull1998value,
  title={Value at risk when daily changes in market variables are not normally distributed},
  author={Hull, John and White, Alan and others},
  journal={Journal of derivatives},
  volume={5},
  pages={9--19},
  year={1998},
  publisher={INSTITUTIONAL INVESTOR INC.}
}

@article{BaroneAdesi1999,
  author  = {Giovanni Barone-Adesi and Kostas Giannopoulos and Les Vosper},
  title   = {{VaR} without Correlations for Portfolios of Derivative Securities},
  journal = {Journal of Futures Markets},
  volume  = {19},
  number  = {5},
  pages   = {583--602},
  year    = {1999}
}

@article{BollerslevWooldridge1992,
  author  = {Bollerslev, Tim and Wooldridge, Jeffrey M.},
  title   = {Quasi-Maximum Likelihood Estimation and Inference in Dynamic Models with Time-Varying Covariances},
  journal = {Econometric Reviews},
  volume  = {11},
  number  = {2},
  pages   = {143--172},
  year    = {1992},
  doi     = {10.1080/07474939208800229}
}

@article{Bollerslev1986,
  author  = {Tim Bollerslev},
  title   = {Generalized Autoregressive Conditional Heteroskedasticity},
  journal = {Journal of Econometrics},
  volume  = {31},
  number  = {3},
  pages   = {307--327},
  year    = {1986}
}

\end{document}